\documentclass[twocolumn, switch]{article}
\usepackage{preprint}
\usepackage{algorithm}
\usepackage{algorithmic}
\usepackage{amsmath, amsthm, amssymb, amsfonts}
\usepackage[numbers,square,sort&compress]{natbib}
\usepackage[utf8]{inputenc}
\usepackage[T1]{fontenc}
\usepackage{textcomp}
\usepackage{graphicx}
\usepackage{xcolor}
\usepackage[colorlinks=true,linkcolor=cyan,urlcolor=blue,citecolor=cyan,anchorcolor=black]{hyperref}
\usepackage{xr-hyper}
\usepackage{booktabs}
\usepackage{nicefrac}
\usepackage{microtype}
\usepackage{lineno}
\usepackage{float}
\usepackage{afterpage}
\usepackage{dblfloatfix}
\usepackage{caption}
\usepackage{multirow}
\usepackage{tabularx}
\usepackage{longtable}
\usepackage{array}
\usepackage[normalem]{ulem}
\usepackage{tcolorbox}
\tcbuselibrary{breakable,skins}
\usepackage{titlesec}
\titlespacing\section{0pt}{12pt plus 3pt minus 3pt}{1pt plus 1pt minus 1pt}
\titlespacing\subsection{0pt}{10pt plus 3pt minus 3pt}{1pt plus 1pt minus 1pt}
\titlespacing\subsubsection{0pt}{8pt plus 3pt minus 3pt}{1pt plus 1pt minus 1pt}

\title{LLM-generated personalized nudges for improving pro-environmental behavior: Field evidence from resource conservation}

\usepackage{titling}
\usepackage{footmisc}
\newcommand{\Author}[2]{\textbf{#1}\textsuperscript{#2}}
\author{
  \Author{Zonghan Li}{1,2} \and
  \Author{Yi Liu}{1,3} \and
  \Author{Chunyan Wang}{1,*} \and
  \Author{Song Tong}{4} \and
  \Author{Kaiping Peng}{5} \and
  \Author{Feng Ji}{2}
}
\date{%
  \small
  \textsuperscript{1}School of Environment, Tsinghua University, Beijing, China\\
  \textsuperscript{2}Department of Applied Psychology and Human Development, University of Toronto, ON, Canada\\
  \textsuperscript{3}State Key Laboratory of Regional Environment and Sustainability, Tsinghua University, Beijing, China\\
  \mbox{\textsuperscript{4}Department of Psychology, Faculty of Arts and Sciences, Beijing Normal University at Zhuhai, Zhuhai, China}\\
  \textsuperscript{5}Department of Psychology and Cognitive Sciences, Tsinghua University, Beijing, China\\
  [4pt]
  \footnotesize \textbf{* Correspondence to:} Chunyan Wang (\texttt{wangchunyan@tsinghua.edu.cn})
}

\begin{document}
\twocolumn[
  \begin{@twocolumnfalse}
    \maketitle
    \thispagestyle{empty}
    \begin{abstract}
    Encouraging pro-environmental behavior remains a major challenge for sustainable cities. Conventional feedback nudges can show individuals how their current behavior compares with environmental goals but often provide limited guidance on what to do differently in daily life. This study examines whether supplementing weekly feedback on participants' behavior with LLM-generated personalized action suggestions improves pro-environmental behavior, using daily electricity and hot-water conservation as a case study. We developed an LLM agent that generated weekly conservation messages from participant profiles, recent consumption records, and prior interaction history, combining a usage report with personalized suggestions, behavioral-change scenarios, and estimated savings. The agent was evaluated in a three-arm randomized field experiment with 233 university residents in Beijing from November 2024 to January 2025. Participants received text-based nudges, image-enhanced nudges, or LLM-generated personalized nudges over five intervention rounds. Daily electricity use and shower hot-water use were measured using dormitory meter readings and billing records. Compared with text-based feedback, LLM-generated personalized nudges reduced electricity consumption by 0.56 kWh per room-day (\emph{p} = 0.014), corresponding to an 18.3 percentage-point higher saving rate. Image-enhanced feedback alone showed no clear improvement. Hot-water savings followed the same direction but were smaller and less precisely estimated (9.8 percentage points, \emph{p} = 0.087). Personalized nudges contained more planning, appliance-specific, and action-oriented language and were associated with more sustained, task-focused engagement. These findings highlight a pathway for integrating generative AI into sustainable urban management.
    \end{abstract}
    \vspace{0.05cm}
    \vspace{0.5cm}
  \end{@twocolumnfalse}
]

\hypertarget{introduction}{%
\section{Introduction}\label{introduction}}

Encouraging residents' pro-environmental behavior is an important component of urban sustainability, a core focus of Sustainable Development Goal 11 \citep{ref1}, and is widely recognized as a key demand-side lever shaping resource use and environmental pressure \citep{ref2}. The costs of pro-environmental behavior are immediate and recurrent, but its returns are delayed and difficult to perceive \citep{ref3}. Even individuals who endorse environmental goals often fail to translate intention into sustained action \citep{ref4}. As cities now concentrate over half of the world's population, projected to reach 68\% by 2050 \citep{ref5}, sustaining pro-environmental behaviors among urban populations has become an urgent but challenging need.

Behavioral nudges are widely used to encourage pro-environmental behavior \citep{ref6}, but their effects are limited in many real-world settings \citep{ref7,ref8}. Nudges are interventions that influence behavior while preserving freedom of choice \citep{ref9}. In the pro-environmental domain, informational nudges are most widely used \citep{ref6,ref10}, as many behaviors (e.g., recycling and household resource conservation) are embedded in routine activities without discrete decision points \citep{ref11,ref12}. Informational nudges provide diagnostic information on behavior, such as feedback, historical comparison, or social comparison, which makes discrepancies between current behavior and desired conservation goals (i.e., behavioral gaps) visible \citep{ref13,ref14,ref15,ref16}. However, identifying a gap does not necessarily specify how individuals should respond to it. Individuals may still need to determine which specific actions to change and how to incorporate them into daily routines. Nudge effects are often modest and inconsistent across studies \citep{ref7}. To address this limitation, nudges have increasingly incorporated prescriptive suggestions on specific actions individuals can take. However, such suggestions are typically standardized for multiple individuals and do not adapt to heterogeneous and changing individual contexts.

Generating prescriptive suggestions for multiple individuals poses a common challenge in personalized interventions across domains. Human professionals can tailor suggestions to individual contexts, such as household energy audits, but this approach does not scale with population size due to considerable marginal cost \citep{ref17,ref18}. Automated systems improve scalability by encoding suggestions in predefined structures, including rule-based matching systems \citep{ref19,ref20} and reinforcement learning approaches that optimize delivery from feedback \citep{ref21,ref22}, but generally require researchers to specify relevant user states, behavioral patterns, as well as knowledge and rules before deployment. Such predefined representations are effective when intervention targets and decision contexts are relatively stable, but become difficult to maintain when individuals differ in their circumstances or when feasible actions change during repeated intervention \citep{ref23,ref24}. Personalization thus faces a persistent trade-off between contextual fit and scalability.

Breaking this trade-off requires mechanisms that can generate suggestions tailored to individuals' evolving circumstances with limited reliance on predefined content or manual creation, and large language models (LLMs) provide one possible approach. An LLM can generate suggestions from individual situations described in natural language \citep{ref25}, allowing generated suggestions to incorporate individual context and be updated across interaction rounds while reducing dependence on fully predefined state-action mappings used in conventional automated systems \citep{ref26}. Because each response is produced through a single model call, this form of generation also reduces marginal cost when scaled to large populations compared with manually created suggestions. This feature has motivated growing interest in applying LLMs to persuasion and behavioral intervention, with studies reporting changes in attitudes \citep{ref27,ref28} and behavioral intentions \citep{ref29,ref30}. However, these outcomes remain at the level of self-report. Given the well-established intention-behavior gap \citep{ref4}, such findings do not establish changes in actual behavior. Where persistence has been examined, intention gains from LLM chatbot conversations did not survive to 45-day follow-up, and no arm increased self-reported uptake \citep{ref31}. The persistence of effects under repeated interaction, where content is continuously updated as new behavioral evidence accumulates, remains untested.

Filling these gaps calls for behavior that can be measured objectively and continuously, under an intervention repeated across rounds. Everyday household electricity and shower hot-water conservation meet these conditions. Both are metered directly, and as curtailment behaviors they demand repeated effort whose costs fall immediately while the benefits stay diffuse and delayed. With modest private stakes but substantial social externalities, they are standard cases in pro-environmental and public-goods research.

This study examines the effect of LLM-generated personalized nudges (abbr. LLM-personalized nudges) on individuals' pro-environmental behavior using everyday electricity and hot-water conservation as a case study. An LLM agent was developed to generate and update nudges from each individual's profile and past interactions. The effectiveness of LLM-personalized nudges was examined against two conventional nudge conditions (i.e., feedback-only) in a five-week randomized controlled trial (RCT) with 233 university residents in Beijing. Temporal patterns and heterogeneity across individuals were explored, and the individual characteristics predicting conservation behaviors were analyzed with machine learning models. The study provides field evidence on LLM-personalized nudges as a means of altering actual pro-environmental behavior in real-world urban settings. The remainder of this paper is organized as follows. Section 2 describes the design of the LLM agent. Section 3 introduces the design of RCT and the calculation of treatment effect. Section 4 explains the analysis methods. Section 5 reports the results. Section 6 summarizes the conclusions and discusses the implications and limitations.

\hypertarget{the-nudge-generation-and-delivery-system}{%
\section{The nudge generation and delivery system}\label{the-nudge-generation-and-delivery-system}}

\hypertarget{the-llm-agent}{%
\subsection{The LLM agent}\label{the-llm-agent}}

\begin{figure*}[!t]
\centering
\includegraphics[width=0.95\textwidth,height=0.78\textheight,keepaspectratio]{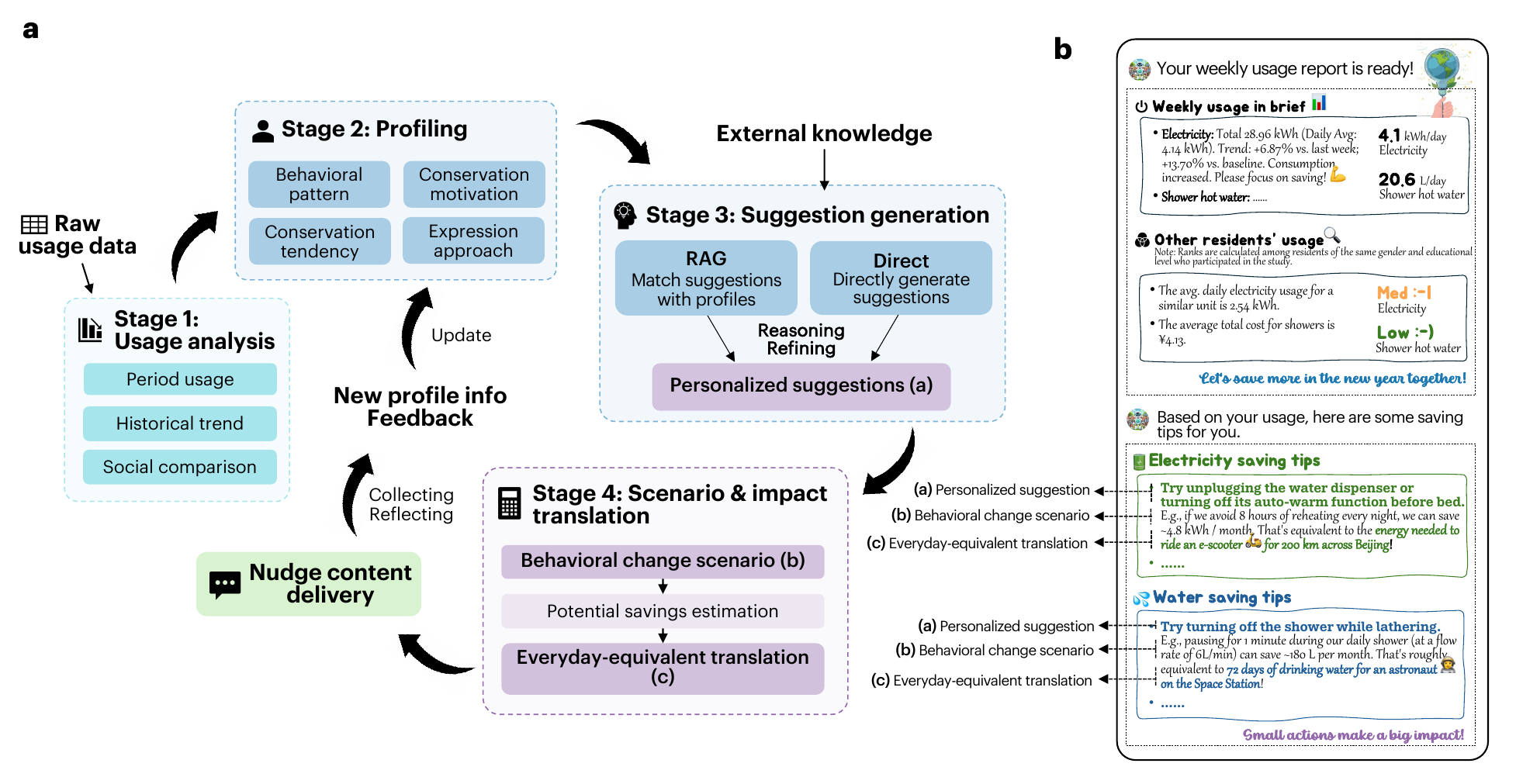}
\caption{\textbf{Design of the LLM agent and an example of the generated nudge content.} \textbf{a.} The agent generates nudge content through a four-stage pipeline: usage analysis (Stage 1), profiling (Stage 2), suggestion generation via retrieval-augmented generation and direct generation (Stage 3), and scenario and everyday-equivalent translation (Stage 4). The profile and interaction record accumulate across rounds, so the cycle repeats each round of nudge, and the content adapts to each participant over the intervention. \textbf{b.} A representative generated nudge, comprising the usage-report backbone and three personalized components: (a) a personalized suggestion, (b) a behavioral-change scenario, and (c) an everyday-equivalent translation of the estimated savings. Consumption values shown are illustrative.}
\label{fig:1}
\end{figure*}

We developed an LLM agent to generate weekly LLM-personalized nudges for electricity and hot-water conservation (Fig.~\ref{fig:1}a). The agent's core function was to analyze the usage data and individual profile, generate personalized conservation suggestions (the prescriptive suggestions described in Section 1) and translate them into participant-specific, context-aware, and action-oriented messages. In each round of nudge, the agent used a fixed prompt structure and updated the individual-level inputs, including individual's baseline profile, sociodemographic information, psychological measures, recent electricity and hot-water records, suggestions delivered in previous rounds, and any comments on earlier nudges. It also maintained a structured participant summary that was refreshed each round, so the weekly message could reflect changes in reported consumption and prior interaction history.

The generated nudge was organized around a feedback-to-action logic shown in Fig.~\ref{fig:1}a. It first made consumption visible through usage analysis, summarizing electricity use, shower hot-water use, recent trends, and comparison with similar participants. It then updated the participant summary through profiling. This profiling step only provided the individual context used in later generation. The agent then produced personalized suggestions by matching the participant summary with relevant records from the suggestion library and rewriting the selected suggestions into plain-language actions. Finally, it generated behavioral-change scenarios and everyday-equivalent translations, placing each suggestion in an everyday situation and expressing the potential savings in approximate and familiar terms.

The agent drew partly on a suggestion library for electricity and hot-water conservation constructed before the trial from 3,219 publicly available documents on everyday electricity and hot-water conservation. The source materials were reorganized into structured records containing the conservation behavior, relevant appliance or setting, conservation strategy, and suggestion text. During generation, the agent retrieved relevant records from this library and adapted them to the participant's profile, recent consumption history, and prior interaction record. It also used the LLM's internal knowledge to generate additional candidate suggestions, pooled them with the retrieved suggestions, and selected the final suggestions based on feasibility, contextual relevance, and expected conservation potential. This combination was intended to keep the suggestions grounded in existing conservation guidance while still allowing them to fit each participant's situation. A random subset of 500 records was manually reviewed to check extraction quality. Details of document collection, record construction, model prompts, model versions, API usage dates, and example records are reported in \textbf{Supplementary Methods S2-S3} and \textbf{Supplementary Tables S7-S8}.

The final message contained four linked elements: a usage report, a personalized suggestion, a behavioral-change scenario, and an everyday-equivalent translation of the estimated savings (Fig.~\ref{fig:1}b). For example, a hot-water suggestion could be expressed as a small change in shower duration, embedded in a familiar shower routine, and translated into approximate monthly savings. The everyday-equivalent translations used uncertainty-marking language to avoid false precision, because their purpose was to make the suggested action interpretable in daily life.

Throughout the generation process, GLM-4-Plus (Zhipu AI) performed suggestion-library retrieval and matching. The OpenAI o1 family was used to generate nudge content. Model versions and API usage dates are listed in \textbf{Supplementary Table S8}. The generation prompts restricted suggestions to low-risk conservation behaviors and prohibited content that could compromise safety, hygiene, or ethical norms. A researcher reviewed all outgoing personalized suggestions against pre-specified safety criteria before delivery, without modifying the content. No suggestions raising such concerns were observed in the delivered messages.

\hypertarget{the-chatbot}{%
\subsection{The chatbot}\label{the-chatbot}}

The agent was delivered through a WeChat chatbot adapted from the open-source ChatGPT-on-WeChat framework. The chatbot appeared to participants as a standard WeChat contact and was used to send weekly nudges, collect daily electricity and hot-water reports, and answer basic conservation-related questions. It integrated the WeChat interface with the LLM APIs and the suggestion library, allowing nudge generation, delivery, data reporting, and participant interaction to occur through the same communication channel.

\hypertarget{experimental-design}{%
\section{Experimental design}\label{experimental-design}}

\begin{figure*}[!t]
\centering
\includegraphics[width=0.95\textwidth,height=0.78\textheight,keepaspectratio]{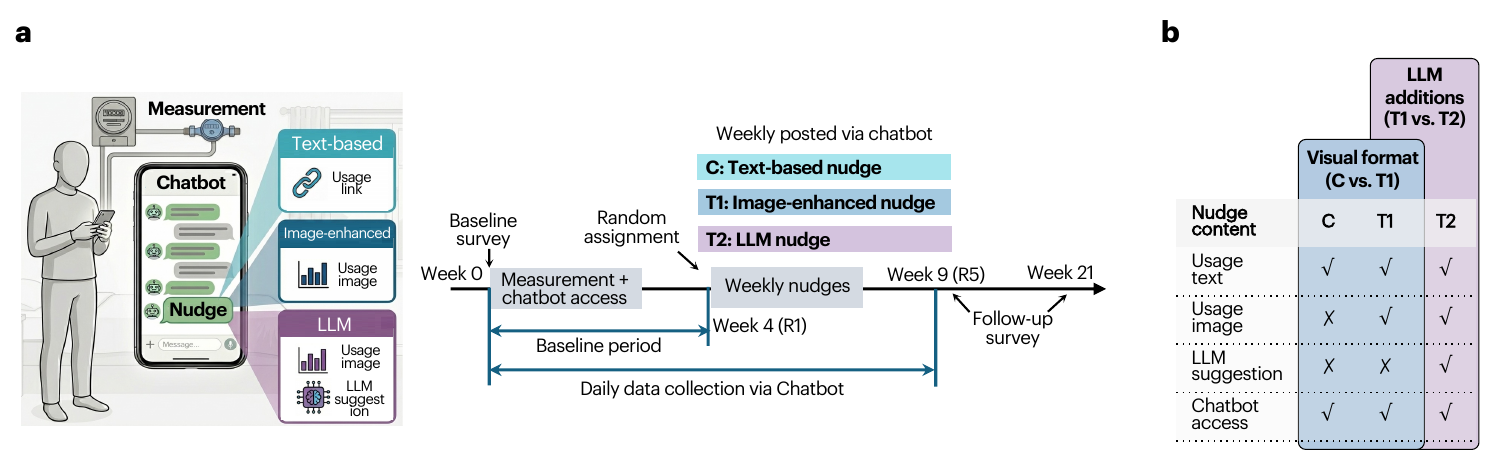}
\caption{\textbf{Study design and intervention components.} \textbf{a.} Overall trial design. \textbf{b.} Decomposition of intervention components across conditions.}
\label{fig:2}
\end{figure*}

To estimate the effect of LLM-personalized nudges on everyday conservation behavior, we conducted a three-arm randomized controlled trial (RCT) among dormitory residents on a university campus in Beijing, China (Fig.~\ref{fig:2}). The arms differed in nudge format and in the presence of LLM-generated personalized components. The primary outcomes were daily electricity use from dormitory meter readings and daily shower hot-water use from billing records. The study ran from November 2024 to January 2025, with a 4-week baseline period followed by a 5-week intervention period (December 12, 2024 to January 16, 2025) in which nudges were delivered weekly for five rounds. We estimated treatment effects by comparing intervention-period consumption across the randomized groups, with baseline consumption and baseline participant characteristics as covariates.

The research was approved by the Tsinghua University Science and Technology Ethics Committee (No.~THU-04-2024-146). All participants provided written informed consent prior to enrollment and received monetary compensation as specified in the consent form. All data were de-identified before analysis and stored in restricted-access environments.

\hypertarget{setting-and-participants}{%
\subsection{Setting and participants}\label{setting-and-participants}}

The trial was conducted on the Tsinghua University campus. Universities house a sizeable share of the urban population, consume substantial water and energy resources, and are widely treated as microcosms of broader urban contexts \citep{ref32,ref33}. They also provide a relatively homogeneous physical and institutional environment (shared infrastructure, tariffs, and facilities), a contextual consistency that reduces unobserved heterogeneity and strengthens internal validity \citep{ref34}.

Participants were recruited through a combination of snowball and purposive sampling, which are often used in field-based behavioral and implementation research \citep{ref35}. Eligible participants were full-time students who lived in on-campus dormitories for the entire study period and could access their dormitory electricity meter readings and shower expenditure records through the university facilities management platform.

A total of 233 students enrolled in the experiment. The sample size was determined by the available participant pool during the recruitment period. Before the experiment, a separate 4-week pilot study was conducted to refine the agent design and the delivery procedures (Fig.~\ref{fig:2}a). To avoid contamination, pilot participants were not eligible for the trial.

\hypertarget{nudge-conditions}{%
\subsection{Nudge conditions}\label{nudge-conditions}}

The experiment compared a text-based nudge condition (control, C), an image-enhanced nudge condition (treatment 1, T1), and an LLM-personalized nudge condition (treatment 2, T2). The same chatbot delivered all three conditions on the same weekly schedule. C and T1, jointly the conventional nudge conditions, contained only the Stage 1 usage report. In C, the weekly chatbot message contained a link, and participants opened the report as a text summary after entering a personal code. In T1, the same report appeared directly in the chat as a concise image that opened with one tap (Fig.~\ref{fig:1}b, upper part). T2 kept the image-report layout of T1 and added the three personalized components produced in Stages 3-4, namely personalized suggestions and their related behavioral-change scenarios and everyday-equivalent translations of the estimated savings (Fig.~\ref{fig:1}b). Comparing T1 with C isolates the effect of presentation and access format. Comparing T2 with T1 isolates the effect of the personalized components (Fig.~\ref{fig:2}b).

\hypertarget{measurement}{%
\subsection{Measurement}\label{measurement}}

The two outcomes were daily electricity use and daily shower hot-water use, and both were obtained by participants from the university facilities management platform. Electricity use was calculated from dormitory meter readings, expressed in kilowatt-hours (kWh). Hot-water use was derived from shower billing records and converted into liters (L) of hot water according to the standard shower tariff. The data were collected each day by the chatbot, which time-stamped every entry.

We assessed the reliability of reported values using redundant records where available. For electricity, a weekly internal validation procedure based on redundant weekly aggregation flagged fewer than 1\% of comparable observations. For hot water, the platform stored hot-water records in both billing and balance formats, so comparable reconciliation was available only for records with balance-based information. In our 4-week pilot study, manually entered values showed no systematic discrepancies from screenshot records. Routine screenshot collection was therefore dropped from the main trial to reduce participant burden.

A baseline survey before randomization measured five pro-environmental psychological constructs and socio-structural characteristics (living budget, gender, and prior experience with electricity bills). The five constructs follow Social Cognitive Theory \citep{ref36} and were validated in our vignette pilot study \citep{ref37}. The peer reference group for the Stage 1 social comparison comprised participants of the same gender and educational level. The survey also contained three implementation-readiness measures (self-reported conservation behavior, perceived room for conservation, and expected helpfulness of the chatbot) for randomization use. We also included three auxiliary measures (post-nudge ratings in T2, chatbot interaction logs, and two follow-up surveys), described in \textbf{Supplementary Methods S4-S6}.

\hypertarget{randomization}{%
\subsection{Randomization}\label{randomization}}

Randomization took place after the baseline period and before the first nudge. To reduce contamination through peer communication, participants who reported enrolling with peers were assigned as a single randomization unit, and all other participants were assigned individually. The three arms contained 77 (C), 78 (T1), and 78 (T2) participants. The three implementation-readiness measures from the baseline survey were used in the covariate-constrained randomization. Covariates were balanced in the sample (\textbf{Supplementary Tables S1-S5}). Details on peer-group size and assignment are reported in \textbf{Supplementary} \textbf{Methods S1}.

Two further safeguards protected the between-group contrast. Recruitment enrolled at most one participant per dormitory room, which kept shared meters and shared living environments out of the comparison. Participants were not informed of the existence of different intervention arms or of the specific content they would receive until the end of the experiment.

\hypertarget{data-cleaning-and-analytic-sample}{%
\subsection{Data cleaning and analytic sample}\label{data-cleaning-and-analytic-sample}}

Data cleaning was conducted while masked to treatment assignment, with the same exclusion rules applied across all experimental arms. Because electricity and hot-water records followed different reporting and validation protocols, data cleaning was conducted separately for the two outcome domains. We first marked invalid data entries that showed non-use, implausible weekly values, or extreme week-to-week instability. We then excluded participants from the outcome-specific analytic sample when more than 40\% of their daily entries during the intervention period were missing or invalid. The final dataset contained 11,146 valid person-day observations, with 169 participants in the electricity sample and 166 in the hot-water sample. Missingness and exclusion rates were comparable across arms (\textbf{Supplementary Tables S1-S5}).

\hypertarget{estimation-and-robustness-checks-for-average-treatment-effects}{%
\subsection{Estimation and robustness checks for average treatment effects}\label{estimation-and-robustness-checks-for-average-treatment-effects}}

We estimated the average treatment effect as the difference in mean daily consumption during the intervention period between nudge conditions, separately for electricity and hot water. Linear models regressed intervention-period consumption on nudge condition. The covariates were baseline consumption, the average pro-environmental psychological score, and socio-structural characteristics. Participants were analyzed by randomized assignment. An omnibus test assessed overall group differences for each behavior, and pre-specified pairwise contrasts among C, T1, and T2 located the source of any difference (\textbf{Supplementary Table S6}). To compare the two behaviors on a common scale, we standardized each outcome within resource type, stacked the electricity and hot-water observations, and estimated pooled models with participant-clustered standard errors, because each participant could contribute one observation per resource. Adjusted saving rates translated these effects into percentage terms. Predicted consumption at sample-mean covariates was expressed as a reduction relative to the sample-average baseline level.

The same specification estimated the treatment effect over time. For each round \emph{k}, mean consumption over rounds 1 to \emph{k} entered the model, and the cumulative saving rates of T1 and T2 were expressed net of the rate in C, which helped account for common temporal trends such as falling winter temperatures.

We assessed robustness using six groups of checks. (1) Randomization-inference checks assessed the sensitivity of the estimates to individual-level independence assumptions and large-sample inference. We re-estimated the primary models with standard errors clustered at the peer-group level, defined by pre-randomization co-participant nominations, and conducted permutation tests that reassigned treatment labels 3,000 times while preserving observed group sizes. (2) Specification checks compared raw unadjusted, baseline-only adjusted, and fully adjusted models. (3) Scale checks compared consumption-level and adjusted-saving-rate parameterizations. (4) Participation-process checks added post-treatment engagement and reporting measures, including reply count, reply length, engagement indicator, and data-report count, as additional covariates to assess sensitivity to differential participation and reporting intensity. (5) Panel checks estimated models with participant fixed effects and week fixed effects spanning baseline and intervention periods, identifying treatment effects from interactions between treatment assignment and the intervention-period indicator. These checks assessed sensitivity to time-invariant individual differences and common weekly shocks. (6) Selection checks computed covariate-adjusted Lee bounds \citep{ref38} to assess sensitivity to selective attrition into the analytic sample. Full robustness specifications are reported in \textbf{Supplementary Table S14}.

\hypertarget{intervention-process-analysis-method}{%
\section{Intervention process analysis method}\label{intervention-process-analysis-method}}

To interpret the behavioral effects of the intervention, we conducted four additional analyses to characterize how the intervention operated and how responses varied across participants. The content analysis examined how the delivered nudges differed across conditions along the textual dimensions examined in this study. The engagement analysis used chatbot logs to measure report opening, text replies, and sustained responsiveness. The heterogeneity analysis used meta-learners and trajectory clustering to describe variation in predicted effects and temporal response patterns. The predictor analysis used machine-learning models to examine measured correlates of intervention-period consumption.

\hypertarget{nudge-content-analysis}{%
\subsection{Nudge content analysis}\label{nudge-content-analysis}}

We analyzed the delivered nudge texts to examine how the LLM-personalized nudges differed from conventional nudges along the dimensions intended by the intervention design. Because C and T1 carried the same information and differed only in presentation format, we pooled them as conventional nudges and compared them with LLM-personalized nudges in T2. Guided by Social Cognitive Theory and an inductive reading of the corpus, we built five keyword dictionaries covering usage-gap feedback, appliance and situational context, planning and action-oriented language, social-norm and identity cues, and encouragement and efficacy content. These categories were used to characterize whether personalized nudges placed greater emphasis on action-oriented language, contextual specificity, social reference, and efficacy-supporting language.

We first compared the prevalence of the five dictionary categories across nudge conditions. As a supplementary check, we estimated a keyword-assisted topic model (keyATM) \citep{ref39} using the same five dictionaries as keyword topics to examine the correspondence between the dictionary patterns and the broader topic structure of the delivered texts. For T2, we further tracked keyword-category changes across rounds to describe how the delivered texts evolved as consumption records, prior suggestions, and participant comments accumulated. The matching procedure, tokenization, and topic-model settings appear in \textbf{Supplementary Method S4} and \textbf{Supplementary Table S11}.

\hypertarget{engagement-and-interaction-analysis}{%
\subsection{Engagement and interaction analysis}\label{engagement-and-interaction-analysis}}

We used chatbot logs to describe participants' exposure and responsiveness to the intervention. Report opening captured exposure to the weekly nudge, while chatbot replies captured active response to the system. We defined engagement as opening at least one weekly report and sending at least one text reply during the intervention period, and computed engagement rates separately for C, T1, and T2.

To describe sustained responsiveness over repeated nudges, we treated the first weekly nudge that received no reply within 48 hours as an event and estimated group-specific Kaplan-Meier curves indexed by round. This analysis measured time to first non-response and was used descriptively to compare the timing of responsiveness across conditions. We also summarized chatbot interaction logs, including session frequency, session duration, usage-query behavior, and multiple replies within 48 hours after nudges, to characterize participants' interaction with the chatbot during the intervention

\hypertarget{heterogeneity-of-nudge-effect}{%
\subsection{Heterogeneity of nudge effect}\label{heterogeneity-of-nudge-effect}}

We examined heterogeneity in nudge effectiveness using two complementary analyses. First, we estimated individual treatment effects (ITEs) with an ensemble of four meta-learners \citep{ref40,ref41}. For each participant, the meta-learners predicted potential outcomes under each nudge condition from baseline characteristics, using the same covariates as the adjusted outcome models. ITEs were calculated as the difference between the predicted outcome under a treatment condition and the predicted outcome under control. Because lower consumption indicates stronger conservation, more negative values correspond to larger predicted reductions in consumption. The ensemble averaged across four learners to reduce dependence on any single model specification.

To describe heterogeneity and profiles associated with larger predicted effects, we compared the ITE distributions of T1 and T2 in the pooled standardized outcome and in each resource domain. We also profiled participants in the top quartile of the predicted T2 effect by comparing their baseline characteristics with those of the remaining T2 participants. The learner definitions, ensemble weighting, cross-fitting procedure, and diagnostic checks are reported in \textbf{Supplementary Methods S8}.

Second, we clustered participants' consumption-change trajectories to identify behavioral archetypes during the intervention. Each trajectory was summarized over early (rounds 1-2), middle (rounds 3-4), and late (round 5) rounds. Participants were first separated by the direction of overall change and then clustered within each direction using agglomerative hierarchical clustering. The clustering used a correlation-based distance, which emphasizes the shape of change over time. The distance measure, clustering procedure, and cluster-selection criteria are reported in \textbf{Supplementary Methods S9} and Fig.~\ref{fig:5}.

\hypertarget{predictors-of-conservation-outcomes}{%
\subsection{Predictors of conservation outcomes}\label{predictors-of-conservation-outcomes}}

We trained machine learning models to examine which measured characteristics were most strongly associated with intervention-period consumption. Average daily electricity use and average daily hot-water use during the intervention were modeled separately using XGBoost \citep{ref42}, which can capture nonlinear associations and interactions among predictors without imposing a linear functional form. The model was selected after comparing multiple linear regression (OLS), decision tree, random forest, XGBoost, and MLP models using cross-validated predictive performance for both outcomes.

Predictors were grouped into baseline consumption, baseline pro-environmental psychological measures, socio-structural characteristics, and intervention-related variables. Intervention-related variables included nudge condition and chatbot interaction measures generated or updated during the intervention, such as chatbot use frequency and average chat length. Because some intervention-related variables were measured after treatment assignment, these models were interpreted as predictive analyses of consumption correlates.

We summarized predictor contributions using normalized gain-based feature importance scores, rescaled to sum to one. These scores indicate predictive importance within the fitted model. To examine changes in predictive patterns over time, we re-estimated the same XGBoost specification for the intervention phases used in the trajectory analysis and compared feature-importance profiles across phases. Full predictor lists, model comparison results, hyperparameter settings, and cross-validated performance metrics are reported in \textbf{Supplementary Tables S12-S13}.

\hypertarget{results}{%
\section{Results}\label{results}}

\hypertarget{behavioral-changes-under-llm-personalized-nudges}{%
\subsection{Behavioral changes under LLM-personalized nudges}\label{behavioral-changes-under-llm-personalized-nudges}}

\hypertarget{average-treatment-effects}{%
\subsubsection{Average treatment effects}\label{average-treatment-effects}}

LLM-personalized nudges (T2) showed lower consumption than both conventional nudge conditions, while image enhancement alone changed little. In the pooled standardized models, consumption in T2 was 0.25 standard deviations (SD) lower than C (\emph{p} = 0.003) and 0.17 SD lower than T1 (\emph{p} = 0.023), with an omnibus group difference at \emph{p} = 0.009. T1 remained close to C (-0.08 SD, \emph{p} = 0.308). The same ordering appeared in both resource domains, with clear patterns for electricity and directionally similar but less precise ones for hot water.

\begin{figure*}[!t]
\centering
\includegraphics[width=0.95\textwidth,height=0.78\textheight,keepaspectratio]{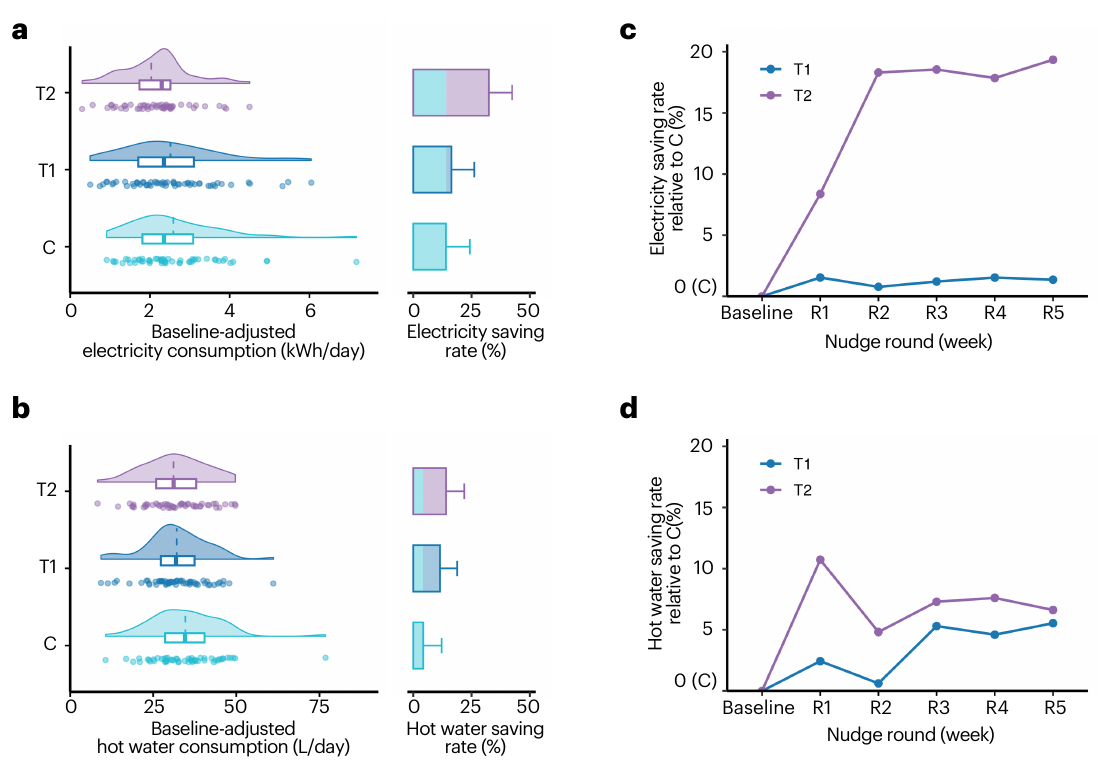}
\caption{\textbf{Conservation outcomes during the intervention.} \textbf{a.} Distribution of individual average daily electricity consumption during the intervention period (left) and group-level saving rates relative to baseline (right) for C, T1, and T2. Bars indicate group means with 95\% confidence intervals. \textbf{b.} Same as panel a, but for hot water. \textbf{c.} Weekly electricity saving rates for T1 and T2 relative to C. \textbf{d.} Same as panel c, but for hot water. Analytic sample sizes were n = 169 for electricity and n = 166 for hot water; group-specific counts are reported in Supplementary Table S1.}
\label{fig:3}
\end{figure*}

For electricity, the saving rate in T2 was more than twice that in C. T2 saved 32.4\% of baseline consumption, compared with 14.1\% in C and 16.0\% in T1 (omnibus \emph{p} = 0.021). In consumption units, T2 used 0.56 kWh per room-day less than C (\emph{p} = 0.014) and 0.49 kWh less than T1 (\emph{p} = 0.023), while T1 differed from C by only 0.07 kWh per room-day (\emph{p} = 0.793; predicted means 2.58, 2.53, and 2.03 kWh per room-day; full estimates in \textbf{Supplementary Table S6}). Over five weeks, the T2-C difference amounted to about 20 kWh per room (Fig.~\ref{fig:3}a). The electricity results therefore show a clear gain from adding personalized content, with little evidence that image enhancement alone reduced consumption.

For hot water, the same ordering appeared with weaker statistical support (Fig.~\ref{fig:3}b). Saving rates were 4.3\% in C, 11.5\% in T1, and 14.1\% in T2 (omnibus \emph{p} = 0.208; predicted means 34.7, 32.1, and 31.1 L per person-day). The T2-C difference was 3.6 L per person-day (\emph{p} = 0.087). The T2-T1 and T1-C contrasts were smaller and imprecise, at 1.0 L (\emph{p} = 0.627) and 2.6 L (\emph{p} = 0.197) per person-day, respectively. In everyday terms, the T2-C point estimate amounted to about 126 L per person over five weeks. The hot-water results therefore support the same directional ordering, but they do not provide a standalone basis for a distinct T2 advantage.

We tested whether these findings held under alternative specifications, and the robustness checks in Section 3.6 did not change the main interpretation. For electricity, T2 continued to show lower consumption than both C and T1 across the randomization-inference, specification, scale, participation-process, panel, and selection checks. For hot water, the checks continued to support a directional T2-C reduction, while the T2-T1 contrast remained inconclusive, including under the selection checks. Full results are reported in \textbf{Supplementary Table S14}.

\hypertarget{temporal-dynamics-across-intervention-rounds}{%
\subsubsection{Temporal dynamics across intervention rounds}\label{temporal-dynamics-across-intervention-rounds}}

Temporal analysis further describes how behavioral responses unfolded across intervention rounds. For electricity, the advantage of T2 appeared early and remained stable after the second round. Round-level saving rates were expressed net of C to account for common temporal trends such as falling winter temperatures. The cumulative saving-rate advantage of T2 over C increased from 8.4 percentage points (p.p.) in round 1 to 18.3 p.p. in round 2, then stayed between 17.9 and 19.3 p.p. for the rest of the intervention (Fig.~\ref{fig:3}c). T1 showed minimal separation from C throughout. Thus, the main separation between LLM-personalized and conventional nudge conditions emerged quickly and then persisted throughout the study period.

For hot water, the T2 advantage appeared earlier but was less stable. T2 exceeded C from round 1, with the advantage narrowing from 10.7 p.p. in round 1 to 4.8 p.p. in round 2 before settling between 6.6 and 7.6 p.p. in rounds 3 to 5 (Fig.~\ref{fig:3}d). T1 accumulated a modest advantage by mid-intervention but remained below T2. Both conditions showed a dip at round 2 before recovering. This pattern is consistent with the average-effect results, as hot-water conservation moved in the same direction as electricity, but the separation was smaller and less sustained.

An exploratory follow-up survey conducted three months after the intervention among C and T2 participants showed little short-term reversal in self-reported conservation for either resource, although this evidence is exploratory because it was self-reported and collected after the main consumption-tracking period (\textbf{Supplementary Method S7}).

\hypertarget{the-intervention-as-delivered-and-received}{%
\subsection{The intervention as delivered and received}\label{the-intervention-as-delivered-and-received}}

\hypertarget{content-of-delivered-nudges}{%
\subsubsection{Content of delivered nudges}\label{content-of-delivered-nudges}}

\begin{figure*}
\centering
\includegraphics[width=0.95\textwidth,height=0.78\textheight,keepaspectratio]{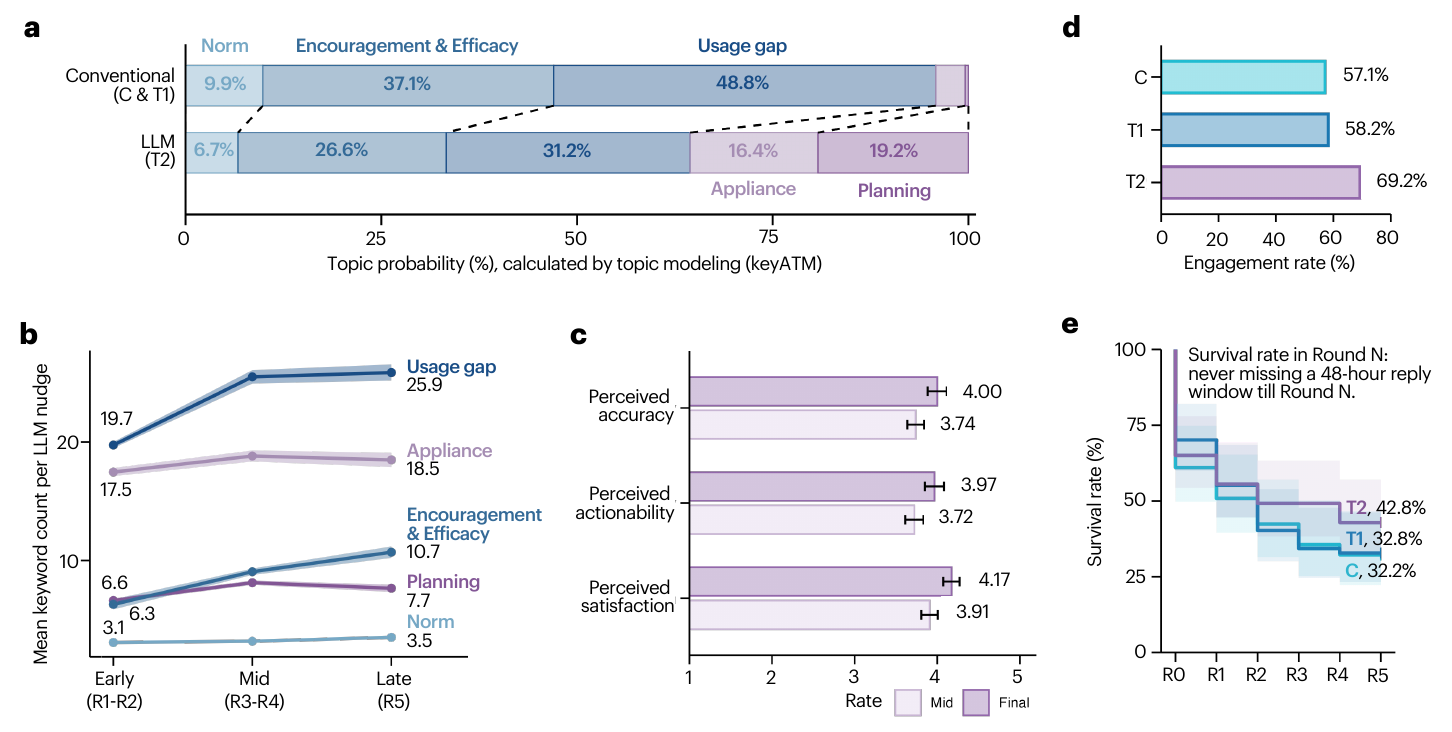}
\caption{\textbf{Content characteristics and participant engagement.} \textbf{a.} Topic composition of delivered nudges estimated by topic modeling, comparing conventional nudges (C \& T1) with LLM-personalized nudges (T2). We merge C and T1 as they share the same information content with differences only in the format. Stacked bars show topic probability (\%). \textbf{b.} Mean keyword counts per LLM-personalized nudge in T2 for five content categories across intervention stages. \textbf{c.} Exploratory post-nudge survey ratings on 5-point scales for perceived accuracy, actionability, and satisfaction in the mid and final rounds among T2 participants. \textbf{d.} Engagement rates for C, T1, and T2. \textbf{e.} Time to the first missed 48-hour reply window following a nudge, with shaded areas indicating 95\% CI.}
\label{fig:4}
\end{figure*}

The delivered texts differed systematically in their content composition, particularly in planning, usage-gap, and appliance-related language. Across all 1,165 delivered nudge messages, LLM-personalized nudges tilted from retrospective description (``what happened'') toward prospective guidance (``what and how to do''). In the topic model (Fig.~\ref{fig:4}a), planning content ran at 19.2\% in LLM-personalized nudges against 0.4\% in conventional nudges, and appliance content at 16.4\% against 3.8\%, while conventional nudges concentrated in usage-gap (48.8\%) and encouragement content (37.1\%). The specificity often took the form of vivid everyday-equivalent translations, including phrases such as ``imagine you'' and ``equivalent to'' (representative messages in \textbf{Supplementary Tables S9-S10}). These patterns indicate a stable divergence in content structure across conditions.

\hypertarget{temporal-adaptation-of-personalized-content}{%
\subsubsection{Temporal adaptation of personalized content}\label{temporal-adaptation-of-personalized-content}}

The content of the LLM-personalized nudges showed systematic variation over rounds. As participant profiles were updated across rounds, usage-gap references increased from 19.7 to 25.9 mentions per nudge, and encouragement content from 6.3 to 10.7 (Fig.~\ref{fig:4}b). Post-intervention survey responses from T2 participants who answered after the last three rounds show parallel shifts in perceived quality, with increases in perceived accuracy (3.74 to 4.00), actionability (3.72 to 3.97), and satisfaction (3.91 to 4.17 on 5-point scales; Fig.~\ref{fig:4}c). These changes indicate that message content evolved across rounds as additional behavioral and interaction information became available, with the same period corresponding to stable treatment effects in electricity consumption (Section 5.1.2).

\hypertarget{exposure-and-interaction}{%
\subsubsection{Exposure and interaction}\label{exposure-and-interaction}}

LLM-personalized nudges reached a larger share of participants and sustained engagement over time. A total of 69.7\% of participants in T2 both opened at least one report and sent at least one reply, compared with 57.1\% in C and 58.2\% in T1 (Fig.~\ref{fig:4}d). Differences in early uptake were modest, but divergence increased after the mid-intervention rounds, with T2 participants maintaining responsiveness for longer. By the final round, the share of participants without any missed 48-hour response window was 10.0 p.p. higher in T2 than in T1 (Fig.~\ref{fig:4}e).

Interaction logs provide finer-grained evidence of this pattern. Participants in T2 engaged in shorter sessions (1.58 vs.~1.90 minutes in T1), were more likely to query their own usage data (9.4\% vs.~5.0\%), and more frequently continued interactions after receiving nudges, as reflected in multiple replies within 48 hours (44.7\% vs.~38.0\%). The interaction profile of T2 participants suggested more task-focused and sustained engagement than that of T1 participants.

\hypertarget{heterogeneity-of-nudge-effects}{%
\subsection{Heterogeneity of nudge effects}\label{heterogeneity-of-nudge-effects}}

\hypertarget{participant-level-heterogeneity}{%
\subsubsection{Participant-level heterogeneity}\label{participant-level-heterogeneity}}

The estimated ITEs separate T2 from T1 more clearly for electricity than for hot water. In the pooled analysis (Fig.~\ref{fig:5}a), T2 shows a more negative mean ITE than T1 (-0.22 SD vs.~-0.08 SD), and a larger share of participants with negative predicted effects (76.7\% vs.~55.2\%). The separation varies by resource domain, with a larger gap in electricity (Fig.~\ref{fig:5}b), where the mean ITE was -0.50 kWh per room-day in T2 against -0.06 in T1. For hot water (Fig.~\ref{fig:5}c), there was a smaller gap, as the means were -2.64 and -1.88 L per person-day, respectively.

\begin{figure*}[!t]
\centering
\includegraphics[width=0.95\textwidth,height=0.78\textheight,keepaspectratio]{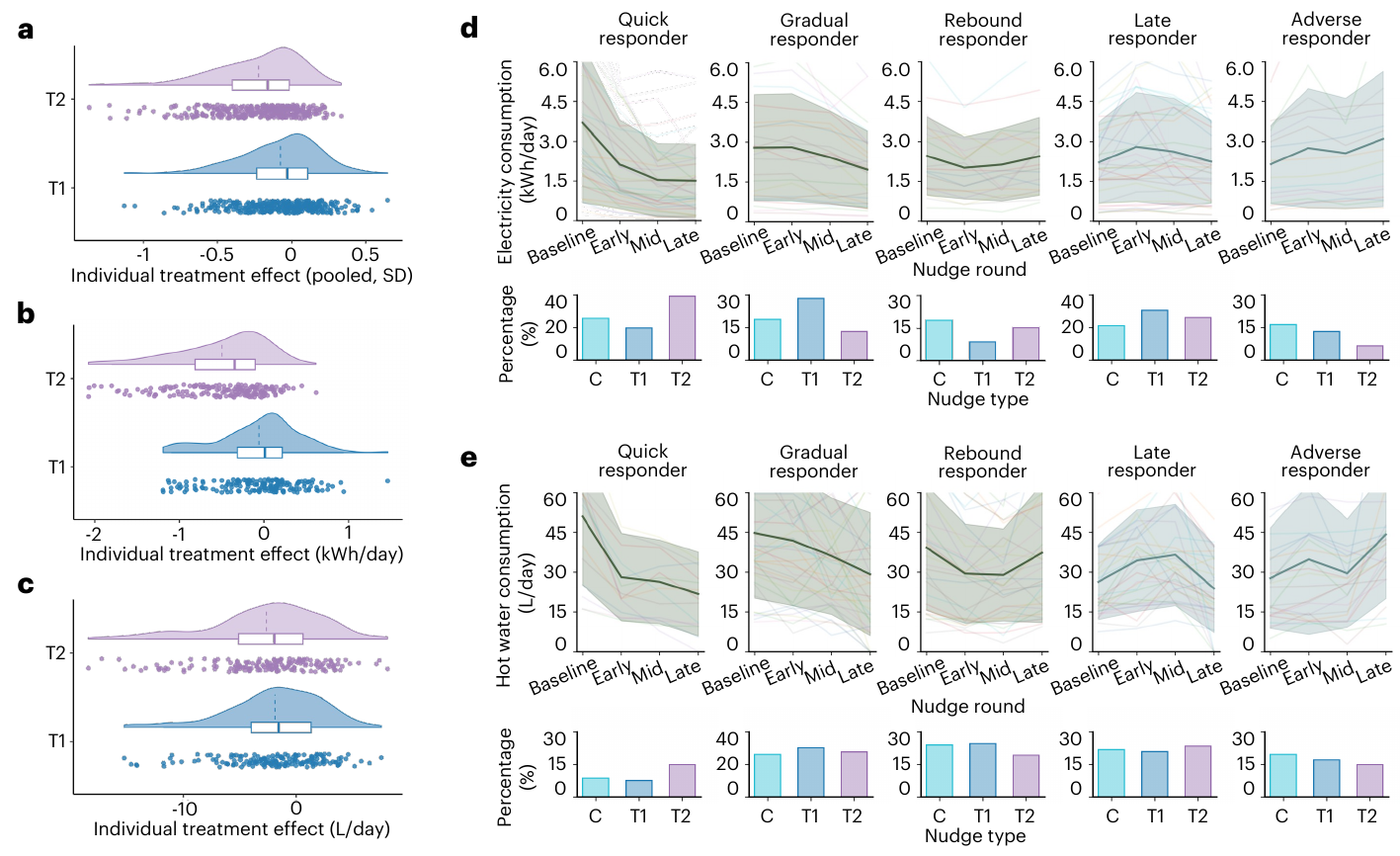}
\caption{\textbf{Heterogeneity of nudge effect.} \textbf{a-c.} Distribution of individual treatment effects (ITEs) estimated via meta-learner for pooled analysis (a), electricity (b) and hot water (c), comparing T1 and T2. Each point represents one participant's ITE relative to the control condition. Distributions are shown as raincloud plots combining kernel density estimates, box plots (median, interquartile range), and individual data points. Dashed lines indicate mean effects. \textbf{d--e.} Behavioral archetypes identified through consumption-change trajectory clustering for electricity (d) and hot water (e). Upper panels show individual consumption trajectories (thin lines) and archetype means (thick lines) across intervention rounds. Lower panels show the percentage of participants in each archetype by nudge condition. Shaded areas denote ±1 s.d. around cluster means.}
\label{fig:5}
\end{figure*}

Within T2, the estimated ITEs were more negative among participants with higher baseline pro-environmental psychological scores and higher living budgets. For electricity, the top ITE quartile had a mean psychological score of 3.77 against 3.51 in the remainder of T2 participants, and living budgets of 2.52 against 1.84 thousand RMB Yuan. Hot water showed a similar pattern (psychological score: 3.73 against 3.58; living budgets: 2.21 against 1.84 thousand RMB Yuan).

\hypertarget{temporal-heterogeneity}{%
\subsubsection{Temporal heterogeneity}\label{temporal-heterogeneity}}

Consumption-change trajectory clustering yielded five behavioral archetypes, including \emph{quick responders} (large early reductions), \emph{gradual responders} (moderate but persistent declines), \emph{rebound responders} (initial reductions followed by partial recovery), \emph{late responders} (early increases followed by partial later reductions), and \emph{adverse responders} (persistent net increases above baseline).

For electricity (Fig.~\ref{fig:5}d), T2 shows a larger share of \emph{quick responders} (39.1\% vs.~19.6\% in T1 and 25.6\% in C) and a smaller share of \emph{gradual responders} (13.0\% vs.~28.3\% in T1 and 18.6\% in C). Among increasers, \emph{adverse responders} are less frequent in T2 (6.5\% vs.~13.0\% in T1 and 16.3\% in C). For hot water (Fig.~\ref{fig:5}e), T2 similarly shows a higher share of \emph{quick responders} (14.9\% vs.~8.7\% in C), a lower share of \emph{rebound responders} (19.1\% vs.~23.9\% in C), and fewer \emph{adverse responders} (14.9\% vs.~19.6\% in C). For both resources, T2 concentrates more participants in early-response trajectories and fewer in unstable or adverse patterns, consistent with the round-level dynamics reported in Section 5.1.2.

\hypertarget{predictors-of-conservation-outcomes-and-their-dynamics}{%
\subsection{Predictors of conservation outcomes and their dynamics}\label{predictors-of-conservation-outcomes-and-their-dynamics}}

\begin{figure*}[!t]
\centering
\includegraphics[width=0.95\textwidth,height=0.78\textheight,keepaspectratio]{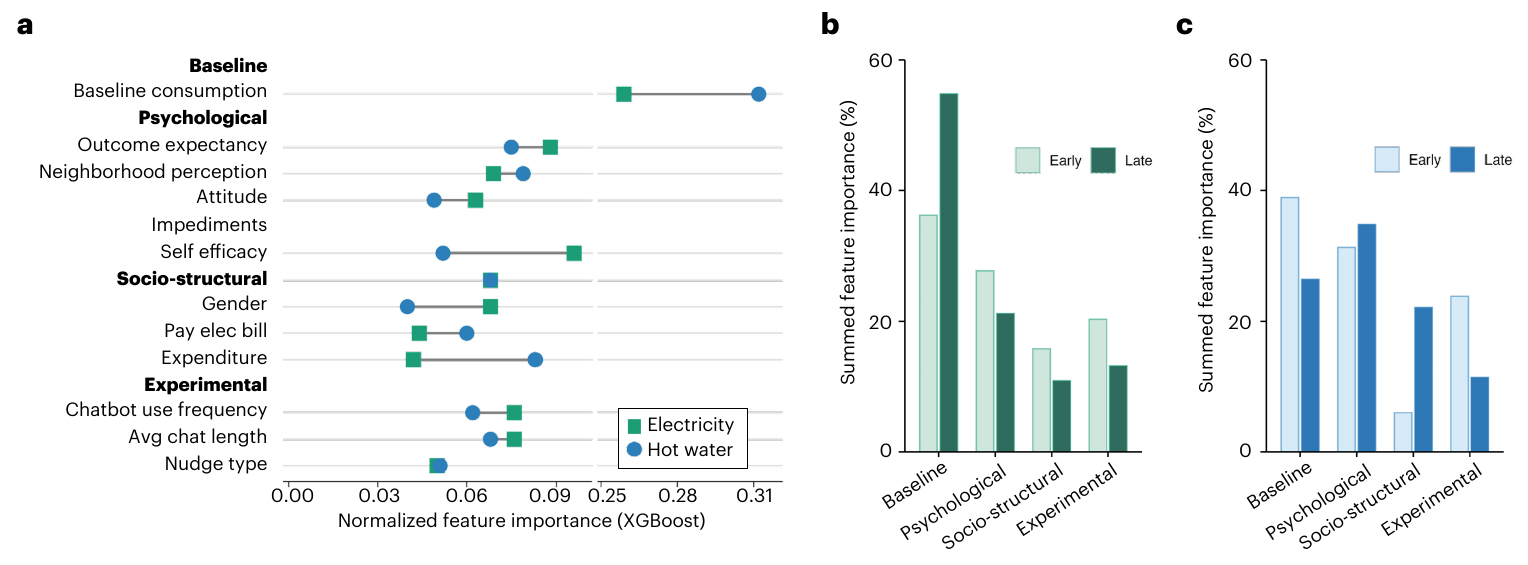}
\caption{\textbf{Predictors of conservation outcomes and their dynamics.} \textbf{a.} Normalized feature importance scores from XGBoost models predicting average daily consumption during the intervention for electricity (green square) and hot water (purple circle). \textbf{b.} Category-level feature importance in early and late intervention stages for electricity. At each stage, a separate XGBoost model was trained using the same feature set. \textbf{c.} Same as panel b, but for hot water.}
\label{fig:6}
\end{figure*}

XGBoost models predicted conservation outcomes (i.e., the intervention-period consumption) from baseline, psychological, socio-structural, and intervention-related variables. For both resources (Fig.~\ref{fig:6}a), baseline consumption led the importance ranking (normalized importance 0.26 for electricity and 0.31 for hot water), so intervention-period consumption remained strongly associated with pre-existing usage patterns and demand levels. Within the psychological measures, outcome expectancy led for electricity (0.09), and self-efficacy (0.07) and neighborhood perception (0.07) led for hot water. Electricity consumption is therefore more associated with expected outcomes, while hot-water consumption is more associated with perceived ability and social context.

Phase-specific models separate the two resources (Fig.~\ref{fig:6}b-c). For electricity, the relative importance of baseline consumption rose from 36.2\% to 54.8\% between early and late rounds. For hot water, it fell from 38.9\% to 26.4\%, while the combined weight of psychological and socio-structural characteristics rose from 37.3\% to 56.9\%. These patterns indicate a divergence in predictive structure over time between the two resources. Late-round hot-water models place relatively more weight on psychological and socio-structural variables and less on baseline usage, alongside the attenuation of the hot-water treatment effect.

\hypertarget{conclusions-and-discussion}{%
\section{Conclusions and discussion}\label{conclusions-and-discussion}}

Personalized behavioral nudging holds promise for encouraging pro-environmental behaviors, yet whether LLMs can meaningfully improve nudge effectiveness remains an open question. We explore this question in a field RCT on daily electricity and hot-water conservation, in which LLM-personalized nudges were added to the same weekly usage report received by all participants. Across both behaviors, the LLM-personalized nudge condition (T2) showed the strongest conservation pattern, while image enhancement alone (T1) did not produce clearly distinguishable outcomes from the text-based nudge condition (C). For electricity conservation, T2 reduced consumption by 0.56 kWh per room-day relative to C (\emph{p} = 0.014), corresponding to an 18.3 p.p. higher saving rate. Hot-water outcomes followed the same direction (9.8 p.p. higher than C) but were less precisely estimated (\emph{p} = 0.087). The LLM-personalized nudge was also associated with higher participant engagement and earlier emergence of conservation gains for electricity.

For research on LLM-based behavioral intervention, the trial adds the step from stated change to actual consumption. The environmental relevance of this literature depends on that step. Reported effects to date concern attitudes, intentions, and self-reported readiness \citep{ref27,ref28,ref29,ref43}, or choices inside constructed tasks \citep{ref30,ref44}, while the benefit these interventions promise comes only when consumption changes. Effects measured at the stated level therefore provide incomplete evidence of environmental impact. Intention predicts action imperfectly \citep{ref45}, and the immediate costs and delayed returns of pro-environmental behavior can even widen the gap further. Sustained human trials with real sustainability outcomes have therefore been called for \citep{ref46}. The present trial extends our earlier vignette evidence, in which LLM-personalized nudges shifted conservation intentions \citep{ref37}, to metered consumption under repeated intervention. Our study therefore supplies the behavioral counterpart to intention-level findings in this literature.

The incremental effect of LLM-personalized nudges can be contextualized by previous attempts to augment feedback with additional informational, normative, or tailored components. In those studies, personalization mainly means inserting household-specific figures into a predefined report format, and they often produce modest or context-dependent gains. For instance, in Mi, et al.~\citep{ref47}, combining normative information with cost-benefit or environmental-contribution feedback increased estimated electricity savings by about 6.1 and 13.9 percentage points relative to the corresponding feedback-only conditions. In Fang, et al.~\citep{ref48}, adding individualized shower energy reports to real-time feedback generated an additional reduction of 0.23 kWh and 3.8 L per shower. Papineau and Rivers \citep{ref49} further showed that households responded more strongly to heat-loss imagery than to a traditional energy report at the same level of estimated dollar savings, 8.1\% versus 1.3\% reduction in gas use per \$100. However, other trials have reported limited or null gains from additional informational, social, or motivational components \citep{ref50,ref51,ref52,ref53}. Against this literature, T2's additional 18.3 percentage-point electricity saving represents a relatively large incremental effect, whereas its 9.8 percentage-point hot-water increment is closer to previous shower and water-conservation interventions.

LLM-personalized nudges may have taken over the action-planning work that follows feedback. Conventional nudges show individuals their past consumption and peer comparisons but often stop at identifying deviations from desired behavior \citep{ref13,ref14}. Individuals may still need to work out their action plan, including which behaviors the gap corresponds to, in which daily situations those behaviors occur, and what to adjust in the next round. In behavioral-change research, action plans and implementation intentions are thought to help convert goals into action because they specify when, where, and how the action takes place \citep{ref54,ref55}. The personalized components in T2 were designed to provide such action-oriented support. They selected action suggestions against the participant's profile, placed them in concrete everyday situations, and updated them each round as new consumption records and prior exchanges accumulated. In the content analysis, the higher share of planning, appliance-specific, and action-oriented language indicates that LLM-personalized nudges went beyond describing consumption and offered prompts closer to action plans. The more sustained and task-focused usage pattern in the interaction logs is consistent with this reading. The incremental effect of T2 over the conventional nudge conditions may therefore reflect a lower cost of moving from a visible usage gap to an executable action.

Human advisors and rule-based systems can likewise turn feedback into action suggestions, and the cost of doing so falls at different points. Human suggestions fit the individual's situation closely, yet their cost accumulates with each household and each round, a scalability problem also discussed in AI-supported coaching work \citep{ref56}. Rule-based systems scale better, but their personalization rests on predefined user categories, behavior patterns, and suggestion mappings, which requires enumerating the possible situations before deployment \citep{ref57}. Individual circumstances keep shifting during an intervention, so that earlier attempts and the previous round's feedback alter what is feasible next. LLM-personalized nudges instead combine an existing suggestion library, the individual's profile, recent consumption, and prior exchanges to recompose the behavioral suggestions for the current situation in each round. This approach allows personalization to be updated during repeated intervention without requiring every possible situation to be specified in advance. This study did not compare against human advisors or rule-based systems directly, and the relative effectiveness of the three remains to be tested. However, as a generating mechanism, LLMs may reduce the dependence of continuous personalization on pre-enumerated rules and hand-authored content, which may therefore make multi-round personalized intervention workable at larger scale in cities.

The difference between electricity and hot-water outcomes suggests that behavioral domains constrain the effectiveness of LLM-personalized nudges. Our results show that electricity savings were larger and more precisely estimated, while hot-water savings were smaller and directional. Electricity conservation admits low-cost marginal or one-time adjustments, such as turning off unused lights \citep{ref58}, while hot-water conservation is tied to bodily comfort and hygiene and carries stronger comfort trade-offs \citep{ref59,ref60}. Electricity savings therefore depend more on the perceived worth of specific actions, and hot-water savings more on the ability to sustain the associated discomfort, which may help explain why electricity conservation aligns more with outcome expectancy and hot-water conservation with self-efficacy. The weaker and less stable hot-water effects suggest that the same personalized nudge mechanism may have less leverage when the target behavior is more embodied, habitual, and costly to sustain. Although the comparison here is limited to two behaviors, the underlying dimensions may vary across other target pro-environmental behaviors.

These behavioral gains should also be considered against the resource and monetary costs of LLM serving. The intervention generated larger measured resource savings than estimated computational resource use, while remaining financially costly at the serving prices during the experiment. Serving one T2 participant for five rounds took about 182,000 tokens and \textyen{}58.2 (about US\$8). Based on published work on the environmental footprints of LLMs \citep{ref61,ref62,ref63}, the computation used 0.05 to 0.3 kWh and about a liter of water per participant. Each T2 participant saved about 20 kWh of electricity and 126 L of hot water over the intervention on average, so the savings can exceed the computational footprint. Utility savings per T2 participant were about US\$2 according to the university's rate, so the nudges were not cost-neutral under the pricing conditions during the experiment. However, model serving costs have declined rapidly, and smaller or more efficient models now reach performance levels similar to earlier high-cost reasoning systems on many application-specific generation tasks. Interventions of this kind therefore become workable in settings that hold continuous consumption data and allow repeated contact, such as university dormitories, residential communities, and utility platforms.

Several limitations of this study point to directions for future work. First, this study tests LLM-personalized nudges on two resource-conservation behaviors. While enabling precise and continuous measurement, these behaviors may not be representative of the broader range of behaviors commonly targeted by nudge interventions (e.g., dietary choices, physical activity, or medication adherence). Whether the observed patterns under LLM-personalized nudges generalize to behaviors with different feedback structures or habit strength remains an open question. Future work should extend this framework to a wider portfolio of target behaviors. Second, our study treats LLM-personalized nudges as a package combining content type, context integration, and iteration. In the current experimental design, we cannot isolate which component contributes to the observed patterns. Although sensitivity analyses adjusting for post-treatment engagement measures suggested that simple engagement differences alone were unlikely to account for the primary results, the current design still cannot separate iterative updating from differences in message richness, actionability, or perceived novelty. Future work should pursue ablation designs that systematically vary these elements to identify active ingredients. Third, the study was conducted in a single winter season at one university using a student sample. This setting enabled continuous behavioral measurement and preserved internal validity \citep{ref33}, and the use of real-world behaviors provides a testbed closer to actual deployment than typical laboratory settings. However, it limits generalizability to other populations and leaves open questions about seasonal variation and the long-term persistence of treatment effects. Additionally, student dormitories in China are more institutionally constrained than the private-kitchen, apartment-style housing common in many Western university settings, which limits the direct comparability of our findings across contexts. The sample size was also modest and constrained by the available participant pool, which may have reduced the precision of between-group comparisons. Larger, adequately powered trials will be needed to estimate treatment effects more precisely.

\section*{Data availability}
In accordance with the ethics approval and data-protection requirements, anonymized data can be provided upon request.

\section*{Code availability}
The code used for this analysis is available at \url{https://github.com/li-zonghan/LLM-personalized-nudge}.

\section*{Conflict of interest}
The authors declare no competing interests.

\section*{Acknowledgements}
This study was supported by the National Natural Science Foundation of China (No.~52470212, 52522010 to C.W.) and the Social Science and Humanities Research Council of Canada (No.~702-2026-1468 to Z.L.). We thank Dr.~Jieyan Liu, Dr.~Yingjie Liu, Wei Wang, and Jialin Geng from the Tsinghua Center for Student Community Management and Service for their assistance with participant recruitment and field implementation. We thank Shiyu Pei and Lujia Bo from the School of Environment, Tsinghua University, and Yaoyao Zhang, Zhaochen Wang, and Zheng Zhang from Xiuzhong College, Tsinghua University, for assistance with data collection. We also thank Siyi Wang from the University of Toronto for assistance with statistical analysis.

\footnotesize
\bibliography{main}
\end{document}